\documentclass{article}%
\pdfoutput=1
\usepackage[top=1.2in,bottom=1.2in,left=1.2in,right=1.2in]{geometry}%
\usepackage{amsmath}%
\usepackage{amsfonts}%
\usepackage{amssymb}%
\usepackage{graphicx}
\providecommand{\U}[1]{\protect\rule{.1in}{.1in}}
\begin{document}

\title{Multiple Time Dimensions\thanks{Thanks to Walter Craig and Jonathan Hackett
for helpful comments.}}
\author{Steven Weinstein\thanks{Email:\ sw@uwaterloo.ca;
sweinstein@perimeterinstitute.ca.}\\{\small University of Waterloo Dept. of Philosophy}\\{\small 200 University Ave W, Waterloo, ON\ N2L 3G1, Canada}\\{\small Perimeter Institute for Theoretical Physics}\\{\small 31 Caroline St N, Waterloo, ON\ N2L 2Y5, Canada} }
\date{}
\maketitle

\begin{abstract}
The possibility of physics in multiple time dimensions is investigated.
Drawing on recent work by Walter Craig and myself \cite{CW08}, I show that,
contrary to conventional wisdom, there is a well-posed initial value
problem---deterministic, stable evolution---for theories in multiple time
dimensions. Though similar in many ways to ordinary, single-time theories,
multi-time theories have some rather intriguing properties which suggest new
directions for the understanding of fundamental physics.

\end{abstract}

\section{Introduction}

The theoretical framework of physics has evolved enormously since the time of
Newton, but one notable invariant, so pervasive as to be effectively
invisible, is the one-dimensionality of time. While time and space have been
amalgamated into a composite known as \emph{spacetime} in the wake of
relativity theory, and while modern superstring theories follow the
Kaluza-Klein theory in postulating more than three \emph{space} dimensions,
time itself has remained one-dimensional.

Indeed, very little work has been devoted to the study of multiple time
dimensions.\footnote{A notable exception is the recent work of Bars
\cite{Bars00}, which however treats the single extra time dimension as
\textquotedblleft gauge\textquotedblright, thus unphysical.} Yet one might
like to know more about physics with multiple times for at least two reasons:

\begin{itemize}
\item It's not at all clear we can be confident that our world has a single
time dimension unless we know what a world with multiple times looks like.
Kant thought such a world was inconceivable. But Kant also thought that space
must be three-dimensional and Euclidean \cite{Kant29}.

\item Problems connected with the interpretation of quantum mechanics, the
construction and interpretation of a quantum theory of gravity, and the origin
of cosmological time asymmetry all suggest the need for a new conceptual framework.
\end{itemize}

\noindent These questions motivate recent work of Walter Craig and myself
\cite{CW08}, work which explores, from a mathematical perspective, the
features one might expect in a theory with multiple time dimensions. The
results are surprising, undermining as they do the conventional wisdom that
such theories are plagued by instabilities \cite{Dor70} or are hopelessly
unpredictive \cite{Teg97}. As I'll show, theories in multiple time dimensions
allow a meaningful sense of determinism, while giving rise to intriguingly
nonlocal constraints on initial data. Furthermore, the way in which these
sorts of constraints arise, from a reconception of the structure of the
spacetime background, suggests heretofore unexplored ways of extending
physical theory.

\section{Physics in one time and many}

\subsection{One time}

The special theory of relativity brought with it the relativity of
simultaneity, which in turn prompted a reimagining of space and time as
\emph{spacetime}, a four-dimensional manifold of points $M$ equipped with a
metric $g$, the latter giving the distance between pairs of nearby points. If
the square of the distance is positive, then the distance is spatial, and if
it's negative, the distance is temporal. We say that the \emph{signature}\ of
the metric on a 4-dimensional manifold $M~$ metric is $(-,+,+,+)$ if three of
the directions are spatial and one is temporal.\footnote{Note that the choice
of sign is a matter of convention; it could just as well be $(-,-,-,+)$.} The
signature of a 5-dimensional metric with four space dimensions and one time
dimension is thus $(-,+,+,+,+)$, whereas a metric with three space and two
time dimensions has signature $(-,-,+,+,+)$. Thus it is straightforward to
characterize spacetimes with any given number of space and time dimensions.

Matter generally takes the form of either particles or fields, though extended
objects such as strings and membranes are also possible. The focus here will
be on fields, in particular the massless scalar field $\phi=\phi(x_{1}%
,x_{2},x_{3},t)$ described by the `wave equation'
\begin{equation}
\left(  \frac{\partial^{2}}{\partial x_{1}^{2}}+\frac{\partial^{2}}{\partial
x_{2}^{2}}+\frac{\partial^{2}}{\partial x_{3}^{2}}\right)  \phi=\frac
{\partial^{2}}{\partial t^{2}}\phi\text{.}\label{Wave}%
\end{equation}
This is a simple equation in three space dimensions and one time dimension
which describes many phenomena of interest, most notably the propagation of
the components of the electromagnetic field (thus the behavior of light).

There is a `well-posed'\ initial value problem for this equation. What this
means is that if we are given sufficient information about the field at a
given time, a stable solution of the equation exists, and it is unique. In
other words, the initial data completely determine the data at all other
times, and do so in such a way that small errors in the specification of the
initial data do not lead to uncontrollable errors in the solution.

In the usual case, in which the initial data lies on a hypersurface of
codimension one (meaning a hypersurface of dimension one less than the total
dimension of spacetime), the initial value problem is called the Cauchy
problem. Because the equation contains only second derivatives, the
appropriate initial data for the Cauchy problem consist of the field and its
first normal derivative at each point. (The `normal' derivative is the
derivative perpendicular to the hypersurface, which is the derivative in the
time direction.) This is given by the functions
\begin{align}
f(x)  & =\phi(x,0)\label{Initial}\\
g(x)  & =\frac{\partial}{\partial t}\phi(x,0)\nonumber
\end{align}
(where $x$ stands for $(x_{1},x_{2},x_{3})$). The statement that the Cauchy
problem for the wave equation is well-posed means that, given appropriately
differentiable functions $f$ and $g$ representing the relevant properties of
the field at some time, a unique, stable solution exists for all times.

\subsection{Many times}

The generalization of the wave equation to a spacetime with $n$ space
dimensions and $m$ time dimensions is the `ultrahyperbolic'\ equation%
\begin{equation}
\left(  \frac{\partial^{2}}{\partial x_{1}^{2}}+...+\frac{\partial^{2}%
}{\partial x_{n}^{2}}\right)  \phi(x,t)=\left(  \frac{\partial^{2}}{\partial
t_{1}^{2}}+...+\frac{\partial^{2}}{\partial t_{m}^{2}}\right)  \phi
(x,t)\label{Ultra}%
\end{equation}
where $x=(x_{1},...,x_{n})$ and $t=(t_{1},...,t_{m})$ and where both $n$ and
$m$ are greater than $1$. Let's now investigate the status of the initial
value \emph{problems} for this equation, where I say \textquotedblleft
problems\textquotedblright\ rather than \textquotedblleft
problem\textquotedblright\ in recognition of the fact that the meaning of
\textquotedblleft initial\textquotedblright\ is up for grabs in the presence
of more than one time dimension.

\subsubsection{$t_{1}=0$:\ The Cauchy problem}

It has long been known (to those who know it) that the ordinary Cauchy problem
for equation (\ref{Ultra})---the initial value problem on a surface of
codimension one, i.e. dimension $n+m-1$--- is not well-posed. It was shown by
Courant \cite{Cou62}, using the mean-value theorem of Asgeirsson, that
solutions of the equation do not exist for arbitrary choices of initial data%
\begin{align}
f(x,t^{\prime})  & =\phi(x,t^{\prime})\label{Initial_multi}\\
g(x,t^{\prime})  & =\frac{\partial}{\partial t_{1}}\phi(x,t^{\prime}%
)\text{,}\nonumber
\end{align}
where {\small \ }$t^{\prime}=(t_{2},...,t_{m})$ and $x=(x_{1},...,x_{n})$ as
before. This is perhaps unsurprising, given that the initial hypersurface is a
`mixed' hypersurface$\footnote{This is also misleadingly called a `timelike'
hypersurface in \cite{Joh91} and a `non-space-like' hypersurface in
\cite{Cou62}.}$, extended not only in $n$ space dimensions but also in $m-1$
time dimensions. Therefore, it is traversed by lightlike lines, so-called
`characteristics' along which one expects disturbances in the field to
propagate.%
%TCIMACRO{\FRAME{dtbpFU}{3.4592in}{1.8939in}{0pt}{\Qcb{Initial data on a mixed
%hypersurface (dashed lines represent lightlike lines).}}{}{initial.png}%
%width 3.4592in;  height 1.8939in;  depth 0pt;  original-width 3.4134in;
%original-height 1.8568in;  cropleft "0";  croptop "1";  cropright "1";
%cropbottom "0";  filename 'Initial.png';file-properties "XNPEU";}}}%
%BeginExpansion
\begin{center}
\includegraphics[
natheight=1.856800in,
natwidth=3.413400in,
height=1.8939in,
width=3.4592in
]%
{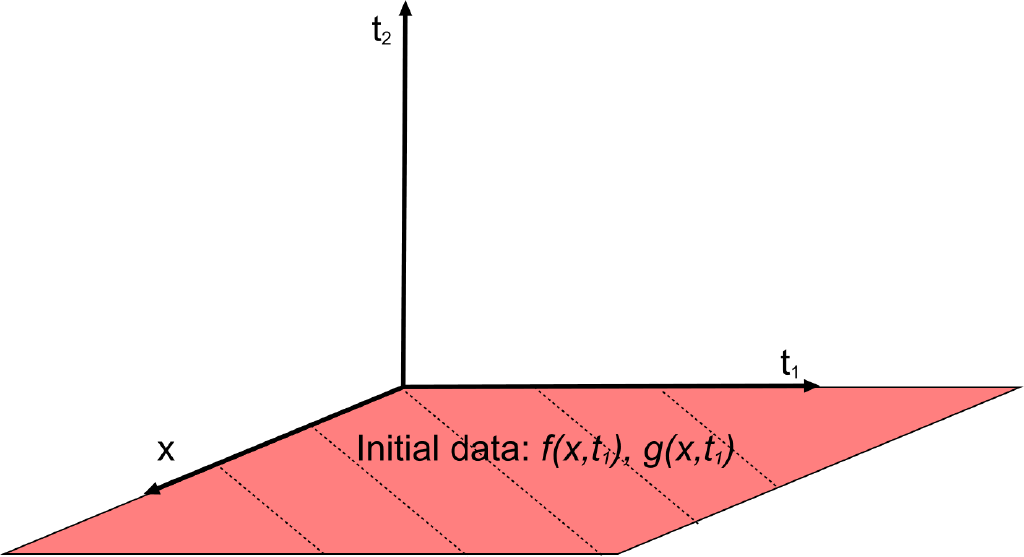}%
\\
Initial data on a mixed hypersurface (dashed lines represent lightlike lines).
\end{center}
%EndExpansion

It might be thought that there is an additional obstacle to well-posedness,
that just as (global) solutions do not exist at all for some initial data,
other data are consistent with \emph{multiple }solutions, conflicting with the
uniqueness requirement. However, this is not the case: the Holmgren-John
uniqueness theorem guarantees that Cauchy data on our mixed hypersurface
uniquely determine the solution everywhere, as long as that initial data is
consistent with \emph{some }solution.\footnote{The original theorem is due to
Holmgren, but is given a more general treatment in \cite{Joh91}.} Indeed, it
tells us that domains of dependence and influence are compact, so that we only
need to know the solution on a compact region $R$ of the Cauchy surface in
order to determine the solution at a given point $E$ off the surface.%
%TCIMACRO{\FRAME{dtbpFU}{3.4592in}{1.8939in}{0pt}{\Qcb{Data in $R$ on a mixed
%hypersurface determines data out to $E$.}}{}{tlike3.png}%
%width 3.4592in;  height 1.8939in;  depth 0pt;  original-width 3.4134in;
%original-height 1.8568in;  cropleft "0";  croptop "1";  cropright "1";
%cropbottom "0";  filename 'Tlike3.png';file-properties "XNPEU";}}}%
%BeginExpansion
\begin{center}
\includegraphics[
natheight=1.856800in,
natwidth=3.413400in,
height=1.8939in,
width=3.4592in
]%
{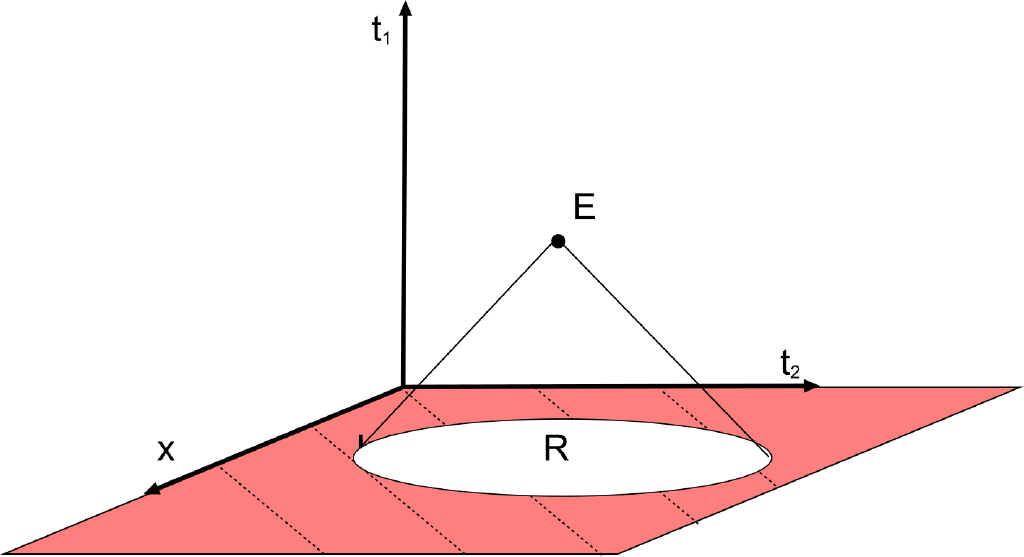}%
\\
Data in $R$ on a mixed hypersurface determines data out to $E$.
\end{center}
%EndExpansion

\noindent This holds true even for mixed surfaces in ordinary spacetime
(replace $x$ with $x_{2}$ $,$ $t_{1}$ with $x_{1}$, and $t_{2}$ with $t$ in
the figure above).

The absence of a well-posed initial value problem is tantamount to a lack of
any sort of practical predictability, and it has been argued that observers
\emph{qua }information processors could not even exist in such a universe, as
they would be unable to engage in any meaningful action in response to
information gleaned about their environment. Thus it has been argued that
universes with more than one time dimension are not meaningful possibilities
\cite{Teg97}. However, recent work of Walter Craig and myself \cite{CW08}
shows that this judgment is too hasty. Imposition of a constraint on the
initial data yields a well--posed Cauchy problem after all.

For those unfamiliar with the notion of a constraint, consider Maxwell's
theory of electromagnetism. The values of the electric and magnetic fields at
a given time uniquely determine the evolution of the field. \ But given
arbitrary initial data, which is to say arbitrary electric and magnetic fields
at some time, a solution to the Maxwell equations of motion will not, in
general, exist. Fortunately, Maxwell's theory comes with two constraints,
Gauss's law for electricity $\nabla\cdot E=0$ and for magnetism $\nabla\cdot
B=0$. Initial $E$ (electric) and $B$ (magnetic) fields satisfying these
constraints \emph{do }give rise to unique global solutions. With the
imposition of the constraints, the initial value problem is indeed well-posed.

The situation turns out to be similar for the ultrahyperbolic equation. There
is a constraint on the initial data such that all and only data satisfying the
constraint lead to a (stable) solution. The constraint is most
straightforwardly specified in terms of the Fourier transforms of the initial
data $\hat{f}(k,\omega^{\prime})=\mathcal{F}(f(x,t^{\prime}))$ and $\hat
{g}(k,\omega^{\prime})=\mathcal{F}(g(x,t^{\prime}))$. It consists simply of
the requirement that the domains of $\hat{f}$ and $\hat{g} $ be restricted to
the region
\begin{equation}
\left\vert k\right\vert ^{2}-\left\vert \omega^{\prime}\right\vert ^{2}%
\geq0\text{.}\label{Constraint}%
\end{equation}
The inverse Fourier transforms
\begin{align*}
f(x,t^{\prime})  & =\mathcal{F}^{-1}\mathcal{(}\hat{f}(k,\omega^{\prime}))\\
g(x,t^{\prime})  & =\mathcal{F}^{-1}\mathcal{(}\hat{g}(k,\omega^{\prime}))
\end{align*}
of such functions then correspond to the allowable sets of initial data. With
the imposition of the constraint (\ref{Constraint}), the problem is well-posed.

The constraint (\ref{Constraint}) has an interesting property: it is nonlocal,
in that it establishes nontrivial correlations between the values of the field
at different points on the hypersurface. Below we have an illustration of this
on a surface spanned by one space and one time dimension: Courant \cite{Cou62}
shows that the field in $R^{\prime}$ uniquely determines the field in $R$.%
%TCIMACRO{\FRAME{dtbpFU}{3.4592in}{1.8939in}{0pt}{\Qcb{$R^{\prime}$ determines
%$R$.}}{}{tlike3d.png}{\special{ language "Scientific Word";  type "GRAPHIC";
%width 3.4592in;  height 1.8939in;  depth 0pt;  original-width 3.4134in;
%original-height 1.8568in;  cropleft "0";  croptop "1";  cropright "1";
%cropbottom "0";  filename 'Tlike3d.png';file-properties "XNPEU";}}}%
%BeginExpansion
\begin{center}
\includegraphics[
natheight=1.856800in,
natwidth=3.413400in,
height=1.8939in,
width=3.4592in
]%
{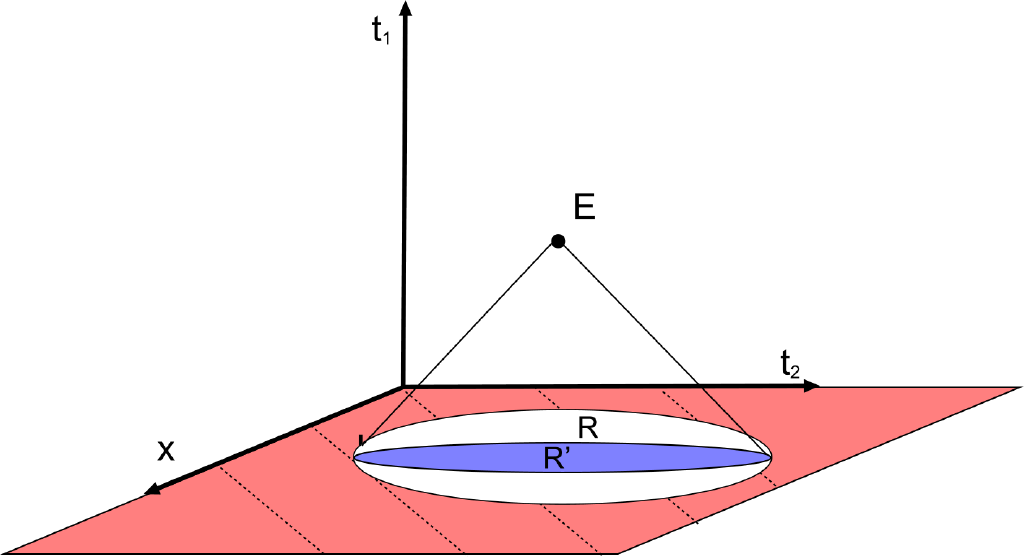}%
\\
$R^{\prime}$ determines $R$.
\end{center}
%EndExpansion
\noindent The nonlocality here is causally benign, since there is no sense in
which changes in one region bring about instantaneous changes in the larger
region; it is a version of what I refer to in \cite{SW08a} as
\textquotedblleft nonlocality without nonlocality.\textquotedblright\ I'll
have more to say about the potential physical significance of such nonlocal
constraints toward the conclusion of this essay.

We've seen, then, that many of the features one physics with one time
dimension remain in the transition to multiple time dimensions, when the
initial value problem is understood as a Cauchy problem.

\begin{itemize}
\item The initial value problem is well-posed, albeit in the company of a
novel, nonlocal constraint.

\item The domains of dependence are compact, as shown by the Holmgren-John
uniqueness theorem.

\item Furthermore, there is a well-defined energy functional---a
Hamiltonian---which is conserved with respect to the chosen time.\footnote{See
Theorem 2 in \cite{CW08}.}
\end{itemize}

\noindent Let's now move on and look at other versions of the initial value
problem, corresponding to other notions of \textquotedblright
initial\textquotedblright\ in the presence of multiple times.

\subsubsection{$t=0$:\ Initial data of higher codimension}

In ordinary physics with a single time dimension, determinism means that the
state of the system at one time determines the state at other times. For a
field theory, the initial data are naturally given on a hypersurface of
codimension one, meaning one fewer dimension than the entire spacetime. The
construction of the Cauchy problem in the previous section simply carries this
over to a spacetime with multiple times, giving data again on a hypersurface
of codimension one, the difference being that the hypersurface is mixed,
rather than purely spacelike. But one might suppose that a more natural way to
give initial data in a theory with multiple times is to give it on a surface
of codimension $m$ (where $m$, again, corresponds to the number of time
dimensions.) In other words, instead of giving data at $t_{1}=0$, we give it
on a purely spacelike hypersurface $t=0$ (which stands for $t_{1}%
=t_{2}=...t_{m}=0$).

Now on the one hand, one might think that the higher codimension problem is
intractable, because there is simply too little information on a surface of
higher codimension to uniquely determine the evolution of the field. After
all, in ordinary physics, we do not expect that giving the values of a field
on a high-codimension surface---e.g. the $x,y$ plane (i.e., the $z=0$
plane)---will suffice to determine the evolution for all times $t$. Not even
close! On the other hand, one might think that the presence of a constraint
could help, since one cannot fill out the rest of the codimension one
hypersurface arbitrarily: one must satisfy the constraint.

It turns out that the constraint is not sufficient to give a unique solution.
Craig and I \cite{CW08} show that the extension of data on $(x,0)=(x,t^{\prime
}=0,t_{1}=0)$ to $(x,t^{\prime},t_{1}=0)$ and then to general $(x,t)$ is
highly nonunique, no matter how much initial data is given on the initial
hypersurface. Even if one gives not just the value of the field and its first
normal derivatives (in the various time directions), but an arbitrary number
of additional normal derivatives, the solution is highly nonunique. So the
higher codimenson problem is, in fact, intractable.

\subsubsection{ $t<\varepsilon$: An \emph{almost}\textbf{-initial} value
problem}

Suppose that instead of giving initial data \emph{on} the hypersurface $t=0$,
one gives it both on the hypersurface and in an arbitrarily small timelike
neighborhood $t<\varepsilon$ of the hypersurface. The result of Courant
\cite{Cou62} discussed earlier has as a consequence that if the data in this
neighborhood are compatible with a solution -- if a solution exists for these
data -- then the solution is uniquely determined in the entire spacetime.
Courant remarks:

\begin{quote}
[W]e are dealing with the remarkable phenomenon of functions which are not
necessarily analytic, yet whose values in an arbitrarily small region
determine the function in a substantially bigger domain. (\cite{Cou62}, 760)
\end{quote}

\noindent So moving from data \emph{on} the higher-codimension hypersurface to
data in the immediate region of the hypersurface changes the nature of the
problem in an essential way. For data on $t=0$, a solution always exists, but
is highly underdetermined, while for data in an arbitrarily small neighborhood
$t<\varepsilon$ of that point, a solution may not exist, but if it does, it is
\emph{unique}. This \textquotedblleft almost-initial-value\textquotedblright%
\ problem is not well-posed, since arbitrarily small changes may take initial
data which are compatible with a solution to initial data which are not
compatible. Nevertheless we have a strong form of determinism.

Note that this phenomenon is not limited to multiple time scenarios. Just as
one can start with almost-initial data on all of space (all points $x$) in the
immediate neighborhood of $t=0$, one can start with data on all of time (all
points $t$) in the neighborhood of some point in space $x=0$. Thus in ordinary
spacetime, with a single time dimension, the field in an arbitrary small
volume of space $x<\varepsilon$ specified at all times $t$ determines the
field \emph{everywhere} in the entire spacetime. Again, one cannot pose
arbitrary data on such a timelike worldtube. But if a solution does exist for
the data given, then it is uniquely determined by the data in the worldtube.
An observer sitting at one point in space for all time would in a sense have
information about the entire, infinite spacetime.

\section{Implications}

We've looked at three sorts of initial value problem for a single, multi-time
theory. The initial value problem on hypersurfaces of higher codimension is
such that one has neither existence nor uniqueness of solution for arbitrary
data, whereas a slight thickening of the higher codimension hypersurfaces
\emph{does} give something closer to a well-posed problem, since solutions are
unique, if they exist at all. When we consider the Cauchy problem, though,
where our initial data is specified on mixed (space and time) hypersurfaces of
codimension one, we finally get something which is remarkably close to
ordinary physics. Here we have a well-posed initial value problem, compact
domains of dependence, and a conserved energy.

Let us focus, then, on the codimension one problem, the Cauchy problem.
Certainly one might wonder about what it means to specify initial data on a
hypersurface which is extended in space and time. But there is no reason
\emph{a priori} why an observer would not just treat the time dimension or
dimensions on the surface as additional spatial dimensions, rewriting for
example
\begin{equation}
\left(  \frac{\partial^{2}}{\partial x_{1}^{2}}+\frac{\partial^{2}}{\partial
x_{2}^{2}}\right)  \phi(x,t)=\left(  \frac{\partial^{2}}{\partial t_{1}^{2}%
}+\frac{\partial^{2}}{\partial t_{2}^{2}}\right)  \phi(x,t)\text{.}%
\label{2+2a}%
\end{equation}
as%
\begin{equation}
\left(  \frac{\partial^{2}}{\partial x_{1}^{2}}+\frac{\partial^{2}}{\partial
x_{2}^{2}}-\frac{\partial^{2}}{\partial t_{2}^{2}}\right)  \phi(x,t)=\frac
{\partial^{2}}{\partial t_{1}^{2}}\phi(x,t)\text{.}\label{2+2b}%
\end{equation}
Leaving aside the question of where and how the observer obtains her time
orientation, let us assume that to some observer, $t_{1}$ looks like time,
whereas $t_{2}$ looks like just another space dimension, albeit one which
enters with a different sign in the equations of motion. From this
perspective, the primary difference between this and the ordinary wave
equation is just the anisotropy of \textquotedblleft space\textquotedblright,
where the scare-quotes indicate that we are referring to the mixed
hypersurface coordinatized by $x_{1},x_{2}$, and $t_{2}$. A further difference
is that the initial value problem is well-posed only for data satisfying the
constraint (\ref{Constraint}), as we've already noted.

From these two features, the presence of a negative sign in the equations of
motion and the consequent requirement of a constraint to ensure the existence
of global (nonsingular) solutions, the observer might infer one of two things:

\begin{enumerate}
\item The world may have begun a finite time ago, as a singularity may lie in
the past, or the end of the world may be nigh, as a singularity may lie in the future.

\item The additional, nonlocal constraint is a feature of the laws of nature,
a feature which guarantees that the evolution is nonsingular, having no
beginning or end.
\end{enumerate}

\noindent I say \textquotedblleft may\textquotedblright\ for (1) since, just
as with the wave equation in ordinary spacetime, the observer cannot predict
arbitrarily far into the future or the past without access to data on the
entire Cauchy surface. On the other hand, the observer might take a different
attitude toward the constraint, and settle on (3) in the belief that nature
abhors a singularity, or in the (related) belief that a finite universe is
absurd. This observer takes the view that the constraint is an additional law
of nature, one which guarantees meaningful nonsingular, global evolution.

One can imagine an argument amongst theorists in this world as to whether the
constraint amounts to an \emph{ad hoc} addition to their laws. One group looks
at the laws and notices that they lead to singularities for arbitrary data and
concludes that there must be an additional law constraining the data. The
other group looks at the laws and from the same evidence concludes that there
may be a singularity in the past or the future: the laws break down.

This is an interesting and suggestive scenario from the perspective of
present-day physics, since the laws that govern the evolution of spacetime do
lead to singularities for arbitrary initial data \cite{HE73}. At the same
time, there is evidence of nonlocality in the large and in the small. In the
large, cosmology is dotted with disturbingly \emph{ad hoc} constraints on the
states of the universe, particularly the low entropy and near-homogeneity of
the early universe. (These are addressed but not resolved by inflation, which
requires its own set of fine-tuned parameters \cite{KKLSS02}.) In the small,
quantum mechanics predicts nonlocal entanglement between the properties of a
given field at various locations in space. It would certainly be worthwhile to
explore whether nonlocal constraints might explain any of these phenomena,
perhaps in conjunction with a modification of the dynamical laws. The sort of
constraint explored in this essay, one arising from the presence of extra time
dimensions, exhibits one sort of nonlocality, but there are other sorts as
well, given by constraints with different functional forms. What they have in
common is that they embody what I have called \textquotedblleft nonlocality
without nonlocality\textquotedblright, meaning nonlocal correlations without
nonlocal causation.\footnote{A more extensive treatment of the relevance of
such constraints to quantum theory may be found in \cite{SW08a}.}

The study of multiple time dimensions here is rather preliminary. I have not
discussed gauge fields or other massive fields, and on a conceptual level I
have not tackled what may be the most difficult question of all, how to
characterize observers and observation in such a theory. What I have shown, I
hope, is that theories with multiple time dimensions are a live conceptual
possibility, and that if nothing else, they serve to stretch our minds as to
what may be physically possible.

\end{document}